\documentclass[%
prl,
showpacs,
numbers, 
amsmath, 
amssymb, 
superscriptaddress
reprint,
]{revtex4-2}

\usepackage{graphicx}
\usepackage{dcolumn}
\usepackage{xcolor}
\usepackage{mathtools}
\usepackage{bm}

\begin{document}

\preprint{APS/123-QED}

\title{Optimal face-to-face coupling for fast self-folding kirigami}

\author{Maks Pecnik Bambic\textsuperscript{1,2,3}, Nuno A. M. Ara\'ujo\textsuperscript{4,5}, Benjamin J. Walker\textsuperscript{6,7}, Duncan R. Hewitt\textsuperscript{7}, Qing Xiang Pei\textsuperscript{2}, Ran Ni\textsuperscript{3}, Giorgio Volpe\textsuperscript{1}} 
\email[Corresponding author: ]{g.volpe@ucl.ac.uk}
\affiliation{\textsuperscript{1}Department of Chemistry, University College London, 20 Gordon Street, WC1H 0AJ London,  United Kingdom \\ \textsuperscript{2}Institute of High Performance Computing, A*STAR, Singapore \\ 
\textsuperscript{3}School of Chemistry, Chemical Engineering and Biotechnology, Nanyang Technological University, 62 Nanyang Drive, 637459, Singapore\\
\textsuperscript{4}Departamento de F\'isica, Faculdade de Ciências, Universidade de Lisboa, 1749-016 Lisboa, Portugal \\ 
\textsuperscript{5}Centro de Física Teórica e Computacional, Faculdade de Ciências, Universidade de Lisboa, 1749-016 Lisboa, Portugal \\ 
\textsuperscript{6}Department of Mathematical Sciences, University of Bath, Claverton Down, Bath, BA2 7AY, United Kingdom \\
\textsuperscript{7}Department of Mathematics, University College London, Gordon Street, London, WC1H 0AY, United Kingdom
}


\begin{abstract}
Kirigami-inspired designs can enable self-folding three-dimensional materials from flat, two-dimensional sheets. Hierarchical designs of connected levels increase the diversity of possible target structures, yet they can lead to longer folding times in the presence of fluctuations. Here, we study the effect of rotational coupling between levels on the self-folding of two-level kirigami designs driven by thermal noise in a fluid. Naturally present due to hydrodynamic resistance, we find that optimization of this coupling as control parameter can significantly improve a structure's self-folding rate and yield.
\end{abstract}


\maketitle

\begin{figure}[t!]
\includegraphics[width = 8.6cm]{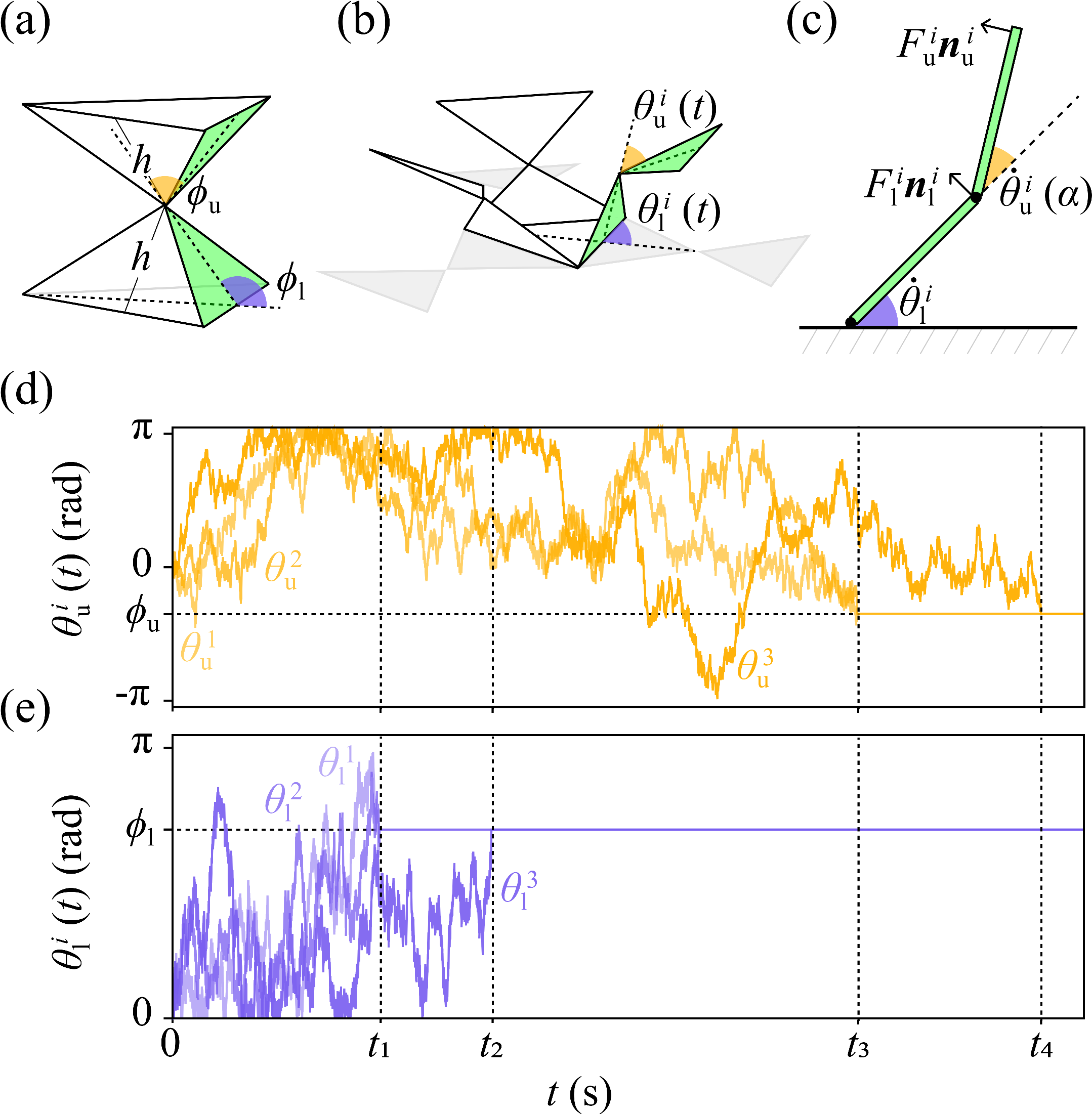}
\caption{\label{fig:fig_1}\textbf{Folding dynamics of a two-level kirigami.} (a) Example of two-level kirigami target structure formed by three sets of two hinged flat sheets (faces) of constant height $h$. The structure is defined by two target angles: $\phi_\textrm{l}$ and $\phi_\textrm{u}$ for the lower and upper levels, respectively. Here, $\phi_\textrm{l} = \frac{\pi}{3} \, {\rm rad}$ and $\phi_\textrm{u} = -\frac{\pi}{6} \, {\rm rad}$ produce an hourglass shape formed by two pyramids touching at their vertices. (b) Two-dimensional template (gray area) for the target structure in (a). The upper faces are hinged to the respective lower ones, which are hinged to the substrate. During folding, the motion of each set of connected faces ($i = 1, \, 2, \, 3$) is described by the angles of the lower and upper levels as a function of time, $\theta_{\rm l}^i(t)$ (defined with respect to the substrate) and $\theta_{\rm u}^i(t)$ (defined with respect to the plane of the lower face).
(c) The folding dynamics can be modeled by considering the overdamped motion of each set $i$ of faces ($\dot \theta_{\rm l}^i$ and $\dot \theta_{\rm u}^i$) under driving forces $F^i_{\rm l}$ and $F^i_{\rm u}$ (here, thermal noise), respectively acting on the lower and upper level (Supplemental Material \cite{SI}). Both forces are directed along the unitary vectors normal to the faces ($\bm{n}_{\rm l}^i$ and $\bm{n}_{\rm u}^i$), causing their rotation around the hinges. Due to the hydrodynamic resistance of the fluid, the motion of the upper faces is naturally coupled to that of the respective lower faces with a dimensionless coupling constant $\alpha <  0$ (here, $\alpha = \alpha_{\rm H} = -5/2$). 
(d-e) Sample trajectories of (d) upper and (e) lower faces. The two horizontal dashed lines indicate the target angles $\phi_{\rm l}$ and $\phi_{\rm u}$ for each level. The vertical dashed lines represent irreversible binding events between two faces: $t_1$ and $t_2$ for the lower level and $t_3$ and $t_4$ for the upper level.
}
\end{figure}

The art of kirigami, where three-dimensional shapes emerge from flat sheets through prescribed cuts and folds, is pursued as a route to design programmable materials and functionalities that can self-fold into a target structure at all scales \cite{Ning2018,Cheng2019,Shanshan2020}. Kirigami-inspired designs are particularly promising due to the easiness of two-dimensional fabrication as well as for their potential for deployability and reconfigurability \cite{Ning2018,Cheng2019,Shanshan2020}. Examples of applications of this strategy include morphable structures and sheets \cite{Kuribayashi2012,Fernandes2012,Guo2013,Fu2018,Paulsen2019}, mechanical actuators \cite{Malachowski2014,Niu2019}, nanocomposites \cite{Erb2013,Shyu2015}, metamaterials \cite{Bertoldi2017,Fang2018,Zhang2022} and soft robots \cite{Wang2018,Rus2018,Miskin2020,Bacchetti2022,dong2022}. 

Due to the broad interest in kirigami designs, the quest to understand the geometrical, topological and mechanical principles behind their folding dynamics has driven a large body of research \cite{Stern2018,Araujo2018,Santangelo2020,Lee2020}, which has also sought to define design rules to optimize them \cite{Pandey2011,Toen2014,Dudte2016,Dieleman2020}.   
At macroscopic scales, as folding is usually driven by stress relaxation \cite{Dudte2016,Santangelo2017,Dieleman2020}, deterministic rules can be identified and relied upon directly \cite{Zhang2015,Dieleman2020,Choi2021,Xiao2022}. However, at microscopic scales, thermal noise can render the folding dynamics a stochastic process \cite{Rothemund2006,Leong2007,Azam2011,Dodd2018,melo2020,Simoes2021}. 

Thermal fluctuations are critical to how microscopic systems explore their configuration space and converge to the desired target structures \cite{Dodd2018,Pandey2011,melo2020,Simoes2021}. Due to these fluctuations, the folding trajectories are stochastic and the final configuration might not coincide with the desired one. The trajectories depend on the choice of the initial template \cite{Pandey2011}, on the properties of the material \cite{Stern2018}, and on the  experimental conditions \cite{Dodd2018}. Folding yield has therefore been considered an important parameter to optimize \cite{Pandey2011,Araujo2018}. Nonetheless, folding time is an equally key parameter beyond yield for real-life applications of microscopic kirigami designs, yet its optimization is much less understood. An accurate prediction of the folding time was possible for simple single level structures (e.g. pyramids) \cite{melo2020,Simoes2021}. However, kirigami designs often present multiple interdependent levels connected via joints, whose folding could be strongly affected by level-to-level correlations, e.g., due to materials or environmental properties \cite{Leong2007,Azam2009,Azam2011,Bao2023}.

Here, we demonstrate how rotational coupling between levels emerges naturally in microscopic hierarchical kirigami templates folding in a fluid. We numerically show how this coupling can be tailored to minimize the folding time and rationalize the choice of the optimal coupling parameter by mapping our results into a first passage problem.

As model target structures, we consider double pyramids fixed on a substrate with three sets of two hinged lateral faces of constant height $h$ (Fig. \ref{fig:fig_1}a-b). Two target angles define the three-dimensional geometry of the structures (Fig. \ref{fig:fig_1}a): $\phi_{\rm l} \in (0, \pi)$ (defined with respect to the substrate) and $\phi_{\rm u} \in (-\pi, \pi)$ (defined with respect to the plane containing the lower face). Based on the choice of angles, the two-dimensional template (gray area, Fig. \ref{fig:fig_1}b) spans from an hourglass shape to a diamond and is obtained by cutting the edges of the lateral faces of the target structure and unfolding them. The upper level faces are hinged to the respective lower level faces, which are in turn tethered to the substrate via a second hinge. Two time-dependent angles, $\theta_{\rm l}(t)$ and $\theta_{\rm u}(t)$, describe the  motion of the two hinges during the folding process driven in the fluid by thermal fluctuations (here, in water at room temperature).

A model for the coupled motion of two faces joined by a hinge can be derived using the approach of ‘resistive-force theory’ \cite{Hancock1953,Gray1955}. Under this approach, the hydrodynamic drag on each face combines with constraints of vanishing net torque and force (owing to the absence of inertia) to yield a coupled equation of motion, while secondary hydrodynamic interactions between the faces are neglected (Supplemental Material for full derivation \cite{SI}). Thus,
\begin{subequations}\label{coupling}
\begin{align}
    \dot{\theta}^i_{\rm u} &= \hat{\alpha}(\theta^i_{\rm u})\dot{\theta}^i_{\rm l} -\frac{3}{C h^2} F^i_{\rm u},\\
    \dot{\theta}^i_{\rm l} &= \hat{\beta}(\theta^i_{\rm u})\dot{\theta}^i_{\rm u} + \frac{6\hat{\beta}(\theta^i_{\rm u})}{\cos(\theta^i_{\rm u}) C h^2} F^i_{\rm l},
\end{align}
\end{subequations}
where $F^i_{\rm l}$ and $F^i_{\rm u}$ are the magnitudes of the forces acting in the normal direction to the lower and upper faces (Fig. \ref{fig:fig_1}c and Fig. S1), respectively, $C<0$ is the normal hydrodynamic resistive force coefficient for each face, which we assume here to be equal (Supplemental Material for the more general case \cite{SI}), and $\hat{\alpha}<0$ and $\hat{\beta}<0$ provide negative feedbacks (i.e. contrary motions) between the faces. 

Equation \ref{coupling} shows how the motion of the two faces joined via a hinge is thus naturally coupled due to the hydrodynamic resistance of the fluid. In fact, for small upper angles $\theta^i_{\rm u}$, the feedback terms in Eq. \ref{coupling} become roughly constant and are dominated by $\hat{\alpha}$, which is $15$ times larger than $\hat{\beta}$ in this limit. Indeed, $\hat{\alpha}$ remains roughly an order of magnitude larger than $\hat{\beta}$ over a wide range of $\theta^i_{\rm u}$ (Supplemental Material \cite{SI}), reflecting the fact that the upper face is much more strongly affected by the motion of the lower than the lower is by the upper.

For microscopic structures, the folding is driven by thermal fluctuations, and the driving forces ($F^i_{\rm l}$ and $F^i_{\rm u}$) change rapidly, being well described by a stochastic process uncorrelated in space and time \cite{Volpe2014}. The variance of this process, and hence the typical rotational diffusion coefficients $D_\theta$ for each face, can thus be derived assuming equipartition (Supplemental Material \cite{SI}). Assuming a hinged circular disc of radius $h/\sqrt{\pi}$, we obtain $D_\theta = \frac{3k_{\rm B} T}{8 \mu h^3}$, where $k_{\rm B}$ is the Boltzmann constant, $T$ the thermostat temperature, and $\mu$ the fluid viscosity (Supplemental Material \cite{SI}). For $h$ being $\mathcal{O}({\rm \mu m})$ to $\mathcal{O}( 10 \, {\rm \mu m})$, $D_\theta$ varies from $\mathcal{O}(10 \, {\rm rad^2 \, s^{-1}})$ to $\mathcal{O}(10^{-3} \, {\rm rad^2 \, s^{-1}})$. 

In order to primarily explore the role of the dominant lower-to-upper coupling of the faces, and taking the sheets as being driven by thermal fluctuations alone, we consider here a reduced model of the form
\begin{subequations}\label{approx}
\begin{align}
    \dot{\theta}^i_{\rm u} & = \alpha\dot{\theta}^i_{\rm l} + \sqrt{2D_{\theta_{\rm u}}}\eta^i_{\theta_{\rm u}},\\
    \dot{\theta}^i_{\rm l} & = \sqrt{2D_{\theta_{\rm l}}}\eta^i_{\theta_{\rm l}},
\end{align}
\end{subequations}
where $D_{\theta_{\rm l}}$ and $D_{\theta_{\rm u}}$ are the rotational diffusion coefficients of the lower and upper face, respectively, and $\eta^i_{\theta_{\rm l}}$ and $\eta^i_{\theta_{\rm l}}$ are independent white noise processes \cite{Volpe2014}. Here, the weaker coupling is neglected and the dominant coupling is parameterized by a single dimensionless constant $\alpha$; a reference value of $\alpha$ is obtained in the limit $\theta^i_{\rm u} \to 0$, which gives $\alpha_{\rm H} \equiv \hat{\alpha}(\theta^i_{\rm u} \to 0) = -5/2$. This follows directly from the model in Eq. \ref{coupling} but, more generally, we leave $\alpha$ as a free parameter to allow for this feedback to be `engineered’ or controlled by other constraints not included in the simple model that underlies Eq. \ref{coupling}. Indeed, even within that modeling framework it is possible to change the strength and direction of coupling by allowing variability in the properties of the two faces and the hinge point between them (Supplemental Material \cite{SI}).

For each set of coupled faces $i$ (Fig.~\ref{fig:fig_1}), we solve these differential equations numerically for $D_{\theta_{\rm l}} = D_{\theta_{\rm u}} = 0.64 \ \textrm{rad}^2 \ \textrm{s}^{-1}$ using a first-order (Euler) integration scheme with a short timestep ($\Delta \tau = 10 \, {\rm \mu s}$) \cite{Volpe2014}. We implement the interaction with the substrate as reflective boundary conditions \cite{Volpe2014} and we correct for collisions among the faces 
by detecting them with the Gilbert-Johnson-Keerthi (GJK) algorithm \cite{Gilbert1988}. We set a cutoff time $\tau_{\rm cut} = 2 \cdot 10^8 \Delta \tau$ for the simulations, when we consider misfolded any structure which is not folded completely. This cutoff time was chosen to guarantee that the rate at which new structures fold is near zero above it. Sample trajectories are shown in Fig. \ref{fig:fig_1}d-e for the faces of the upper and lower levels when $\alpha = \alpha_{\rm H}$, and show how the system converges to the target structure through a series of four irreversible binding events between faces of the lower level first followed by the upper level, where each event is defined by two faces being at the respective target angle ($\pm \pi / 180$) at the same time.

\begin{figure}
\includegraphics[width = 8.6cm]{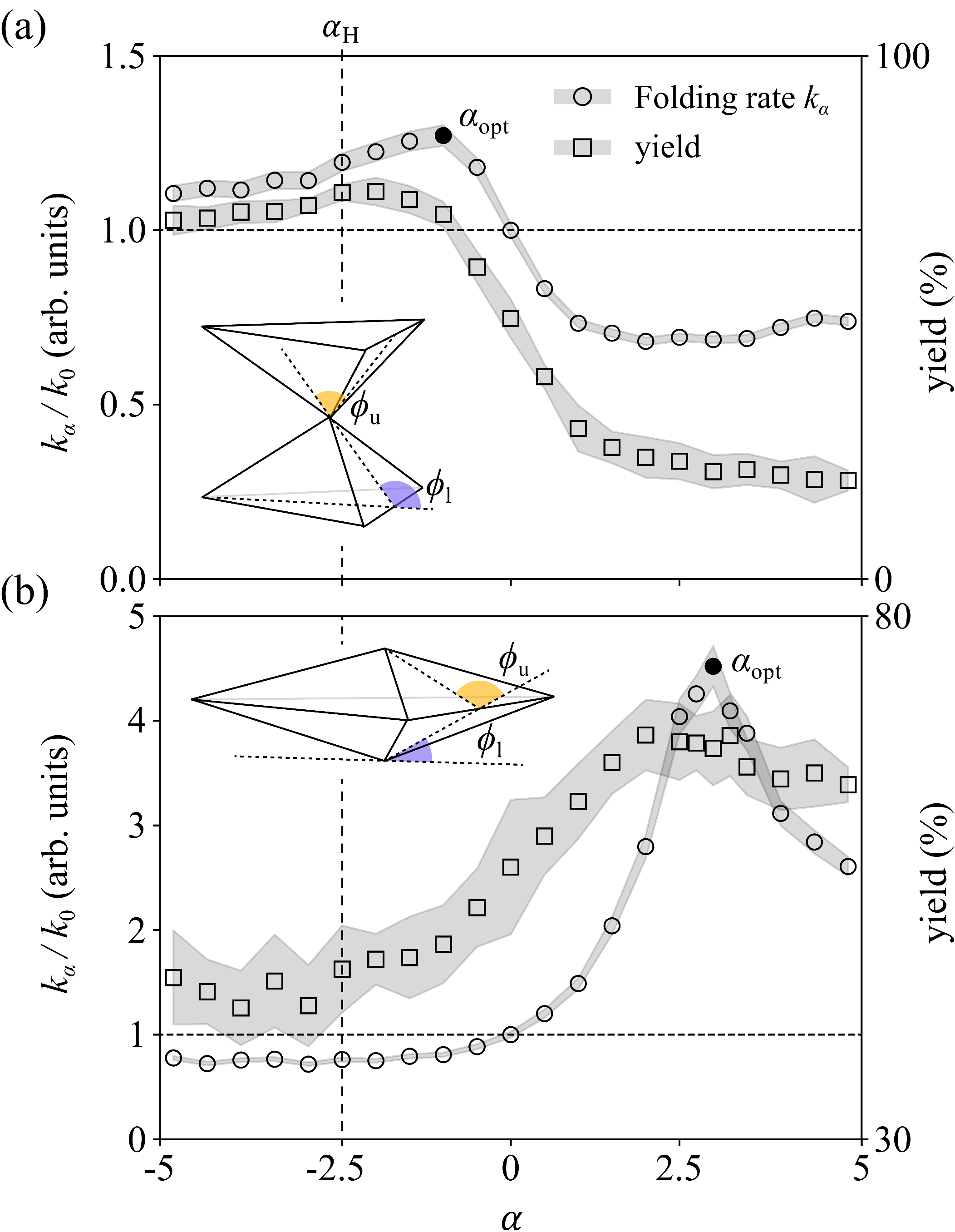}
\caption{\label{fig:fig_2}\textbf{Kirigami folding dependence on the coupling parameter $\alpha$.} 
(a-b) Folding rate  $k_{\alpha}$ (circles) and yield (squares) as a function of coupling parameter $\alpha$ for two exemplary kirigami structures: (a) an hourglass (as in Fig. \ref{fig:fig_1}) and (b) a diamond ($\phi_{\textrm{l}} = 0.61$ rad, $\phi_{\textrm{u}} = 1.92$ rad). 
Depending on the geometry of the target structure, (a) negative or (b) positive values of $\alpha$ can lead to optimal folding ($\alpha_{\rm opt}$, filled circles). 
The vertical dashed lines show $\alpha_\textrm{H}$ for reference. Folding rates are normalized to $k_{0}$ ($k_{\alpha}$ for $\alpha = 0$). The shaded areas represent one standard error around the average values from $5000$ folding events, each lasting up to the cutoff time $\tau_{\rm cut}$. 
}
\end{figure}

\begin{figure}
\includegraphics[width = 8.6cm]{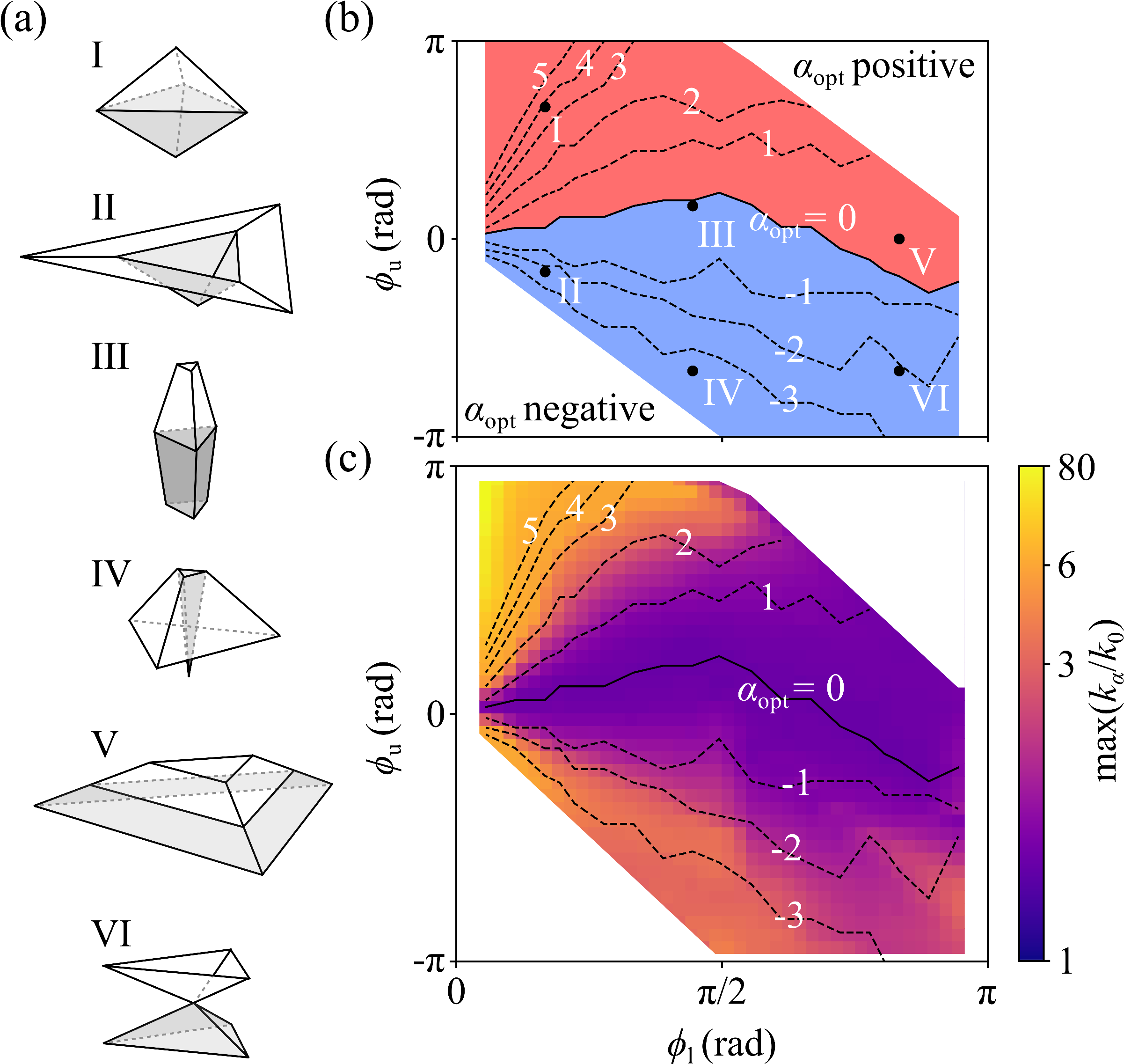}
\caption{\label{fig:fig_3}
\textbf{Optimal coupling for different kirigami target structures.} (a) Examples of different kirigami target structures with constant face height $h$ obtained by varying the target angles $\phi_{\textrm{l}}$ and $\phi_{\textrm{u}}$ (Fig. \ref{fig:fig_1}a). As $\phi_{\textrm{l}}$ increases from acute to obtuse, the target structure of the lower level (gray faces) transitions from an inverted pyramid (I, II, IV) as in Fig. \ref{fig:fig_2}b to a regular pyramid (VI) as in Fig. \ref{fig:fig_2}a via truncated pyramids (III, V). 
For $\phi_\textrm{u}$ going from negative to positive values, the upper level goes from an umbrella shape, either convex (IV) or concave (VI), to a pyramid (I, III, V), truncated or not, through a flat plane (II). 
(b) Phase diagram of the optimal coupling parameter $\alpha_{\rm opt}$ for the different structures. The colour indicates whether a negative (blue) or positive (red) $\alpha$ is optimal (Fig. S4a). (c) Highest folding rate ${\rm max}(k_{\alpha})$ at any $\alpha$ for different target structures. Folding rates are normalized to $k_0$. In (b-c), the black dashed isolines show the optimal values of $\alpha$ that lead to the highest folding rates and the black solid lines highlight structures whose folding is optimal in the absence of coupling ($\alpha = 0$). Each data point is obtained as an average of 5000 folding events.
}
\end{figure}
 
Figure \ref{fig:fig_2} shows that there is scope to engineer the coupling to optimize the folding process by tuning the value -- and sign -- of the coupling parameter $\alpha$. This could be achieved by, e.g., engineering the hydrodynamic resistance of the sheets or exploiting active or driven hinges (Supplemental Material \cite{SI}). These observations are qualitatively independent of the exact choice of the diffusion coefficient (Fig. S3). For example, Fig. \ref{fig:fig_2}a shows that, for the structure in Fig. \ref{fig:fig_1} with an obtuse lower level's target angle and a negative upper level's target angle, the folding rate is enhanced by a factor of $\approx 1.3$ at a slightly negative $\alpha$ ($\alpha = -1$) when compared to the rate without coupling ($\alpha = 0$) and marginally (by a factor of $\approx 1.1$) when compared to the rate at $\alpha_{\rm H}$. 
Instead, for a different target structure such as a diamond with an acute lower level's target angle and a positive upper level's target angle (Fig. \ref{fig:fig_2}b), positive $\alpha$ values can enhance folding by a factor up to $\approx 6$ when compared to the rate at $\alpha_{\rm H}$ and up to $\approx 4.5$ when compared to the no-coupling case. Interestingly, for both structures, the $\alpha$ value that optimizes the rate ($\alpha_{\textrm{opt}}$) is closely related to the value that optimizes yield (defined as the percentage of fully folded structures within the cutoff time $\tau_{\rm cut}$), with a yield of $\approx 70 \%$ (against the highest yield of $\approx 74 \%$) and $\approx 61 \%$ (against the highest yield of $\approx 62 \%$) at $\alpha_{\rm opt}$ in Fig. \ref{fig:fig_2}a-b, respectively. 

Figure \ref{fig:fig_2} suggests that, by correctly designing the coupling between faces, we can optimize both folding rate and yield of the kirigami at once. This observation can be generalized to a variety of structures obtained by varying the two target angles $\phi_{\rm l}$ and $\phi_{\rm u}$ (Figs. \ref{fig:fig_3} and S4). Figure \ref{fig:fig_3}a shows some examples of structures. To a first analysis, the value of $\phi_{\rm u}$ is the main decisive factor discerning whether a negative (for $\phi_{\rm u}<0$) or positive $\alpha$ (for $\phi_{\rm u}>0$) optimizes the structure folding rate (Fig. \ref{fig:fig_3}b). In general, structures with upper levels opening away from the center (e.g. II, IV and VI in Fig. \ref{fig:fig_3}b) tend to benefit from a negative coupling to their respective lower levels, while structures whose upper levels point towards their centers (e.g. I in Fig. \ref{fig:fig_3}b) tend to fold faster with a positive coupling. The absence of coupling instead tends to be optimal for structures (e.g. III in Fig. \ref{fig:fig_3}b and Fig. S4a) where the lower and upper levels roughly lie on the same plane ($\phi_{\rm u} \approx 0$). In fact, Fig. \ref{fig:fig_3}c shows that the benefits in folding rate due to the presence of coupling between levels increases the further a target structure is from this condition, as $k_\alpha \approx k_0$ around $\phi_{\rm u} \approx 0$. This result highlights the importance of engineering the coupling parameter even more, as the folding of target structures in this range of relatively smaller upper angles (Fig. S2) is otherwise inhibited by sub-optimal coupling due to the natural hydrodynamic resistance of the hinged faces (Supplemental Material \cite{SI}). The range of $\phi_{\rm u}$ values where no coupling is advantageous broadens asymmetrically towards positive $\alpha$ values as $\phi_{\rm l}$ increases towards $\pi$. In these situations, stronger coupling tends to push the upper level's faces against the substrate, thus slowing down the convergence to the target structure. As already noted for two exemplary target structures (Fig. \ref{fig:fig_2}), the value of $\alpha$ that optimizes folding rate also roughly optimizes yield (Fig. S4b-d). In fact, for most structures, the ratio between the yield achieved at $\alpha_{\rm opt}$ and the maximum yield at any $\alpha$ (${\rm max(yield)}$, Fig. S4b) is close to one (Fig. S4c), and the distance between these two values of $\alpha$ is often close to zero ($|\alpha_{\rm max(yield)}-\alpha_{\rm opt}| \approx 0$, Fig. S4d). Larger separations between these two $\alpha$ values are possible but often coincide with regions where the yield is relatively insensitive to the exact $\alpha$ value (Fig. S4c) or where advantages over no coupling are negligible (Fig. \ref{fig:fig_3}c).

\begin{figure}
\includegraphics[width = 8.6cm]{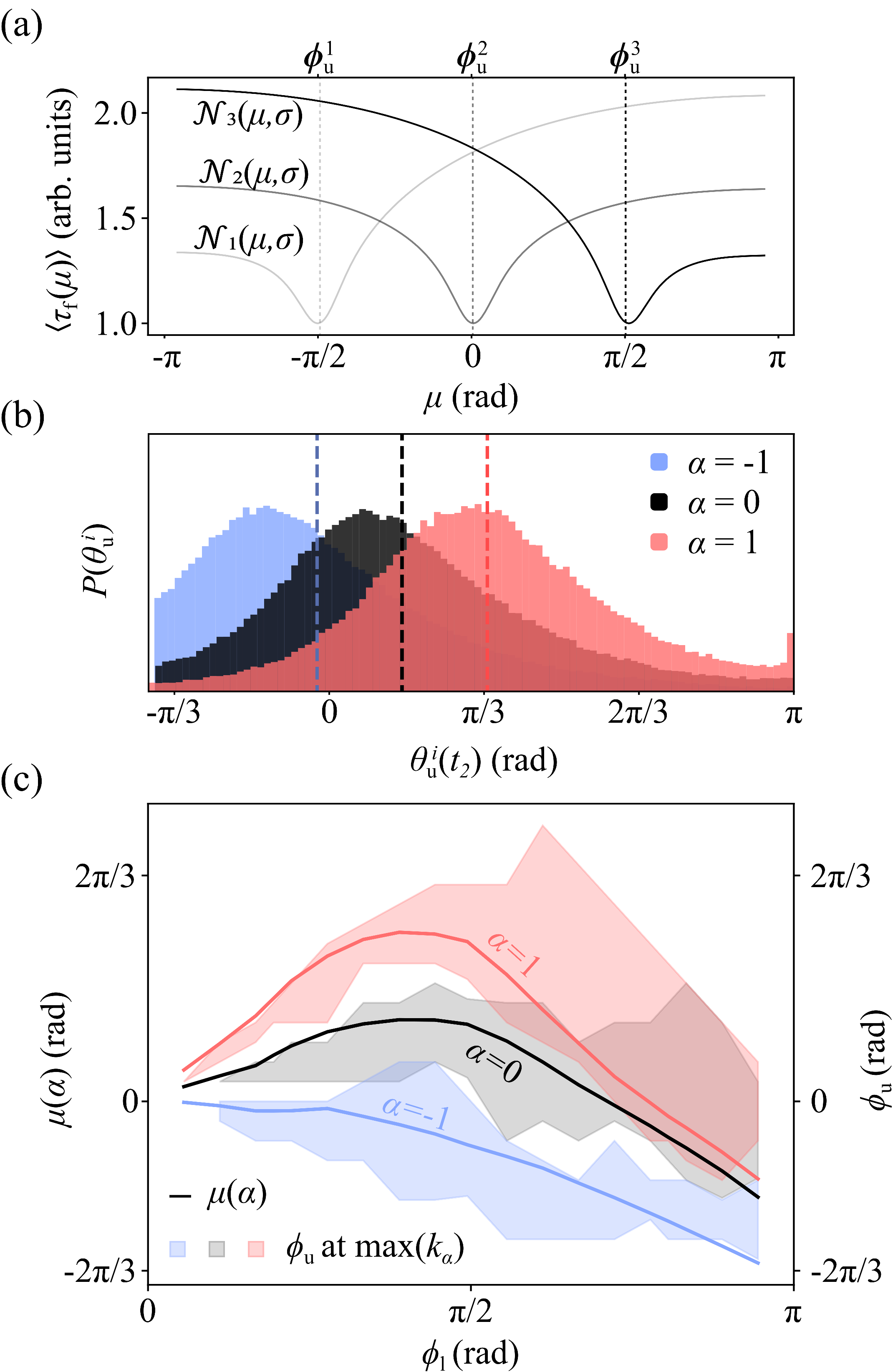}
\caption{\label{fig:fig_4}\textbf{Design rule for optimal coupling.} (a) Expected mean first passage time $\langle{\tau_{\rm f}(\mu)}\rangle$ of two upper level's faces to a target angle $\phi_{\rm u}^n$ (with $n = 1, \, 2, \, 3$) at the closing time of the lower level $t_2$ as a function of the mean $\mu$ of the distribution of the faces' starting position (Supplemental Material \cite{SI}). Here, the starting positions of each face are drawn from a one-dimensional Gaussian distribution $\mathcal{N}_n(\mu, \sigma)$ with varying mean $\mu$, fixed variance $\sigma$ and reflected at the boundaries of the possible range of angles (Supplemental Material \cite{SI}). Three cases are shown: $\phi_{\rm u}^1 = -\frac{\pi}{2}$, $\phi_{\rm u}^2 = \, 0$ and $\phi_{\rm u}^3 = \, \frac{\pi}{2}$. The minimum value of $\langle{\tau_{\rm f}(\mu)}\rangle$ occurs when $\mu \approx \phi_{\rm u}^n$, validating the numerical results in (b-c) analytically. (b) Examples of probability distributions $P(\theta^i_{\rm u})$ of the angles $\theta^i_{\rm u}$ of the upper level's faces at the closing time of the lower level $t_2$ for the structure of Fig. \ref{fig:fig_1} for $\alpha = -1$, $\alpha = \, 0$ and $\alpha = \, 1$. The vertical dashed lines represent the means $\mu(\alpha)=\langle{\theta^i_{\rm u}(t_2)\rangle_{\alpha}}$ of the same-color distributions. Each distribution is obtained from 50000 structures. (c) For a given value of $\alpha$ and lower level's target angle $\phi_{\rm l}$ (see also Fig. S5), the means of these distributions (solid lines) are well captured by the range of possible structures with the highest folding rate at that $\alpha$ value (Fig. \ref{fig:fig_3}c) once the numerical uncertainty in determining ${\rm max}(k_\alpha)$ is accounted for (shaded areas). For a given structure, this uncertainty around the position of ${\rm max}(k_\alpha)$ is calculated taking the largest between the discretization step in $\alpha$ and the range of data points around the peak falling within the standard error.    
}
\end{figure}

To rationalize these observations on the emergence of an optimal coupling parameter $\alpha$ and to provide a design rule for its choice, we can map our results into a first passage problem to a target (Supplemental Material \cite{SI}) \cite{Giuggioli2020}. This choice is justified by recent work that has demonstrated how, for single-level pyramids with less than five faces, the folding time is dominated by the closing of the first pair of faces -- event that can be described as a first passage problem in a two dimensional random walk~\cite{melo2020}. Thus, the total folding time for single-level structures depends on the initial location of its faces \cite{melo2020}. Similarly, for two-level structures, if we assume that the lower level folds much faster than the upper one, the closing of the first pair of upper level's faces after the lower level has closed completely ($t_3$ in Fig. \ref{fig:fig_1}) is the leading event that dominates the folding of most structures. The optimal value of $\alpha$ should then be the one that leads to the distribution of angles in the upper level when the lower level has closed ($t_2$ in Fig. \ref{fig:fig_1}) that minimizes $t_3$ (i.e. the time for the first two faces of the upper level to reach their target angle $\phi_{\rm u}$).  
Starting from a location drawn from a one-dimensional Gaussian distribution $\mathcal{N}(\mu, \sigma)$ with varying mean $\mu(\alpha)$ and fixed variance $\sigma$ and reflected at the boundaries to ensure its proper normalization (Supplemental Material \cite{SI}), the mean first passage time to a trap at location $\pmb{\phi} = (\phi_{\rm u}, \phi_{\rm u})$ of a two-dimensional random walk on a lattice with reflective boundaries can be calculated as \cite{Giuggioli2020}
\begin{equation}
    \langle \tau_\textrm{f}(\mu)\rangle = \iint T_{\bm{\theta}_0\rightarrow \pmb{\phi}} P(\bm{\theta}_0) d\bm{\theta}_0 \, ,
\end{equation}
where $T_{\bm{\theta}_0\rightarrow \pmb{\phi}}$ is the first passage time to the trap starting from position $\bm{\theta}_0 = (\theta_{01}, \theta_{02})$ on the lattice with probability $P(\bm{\theta}_0) = \mathcal{N}(\mu, \sigma)_{|\theta_{01}}\mathcal{N}(\mu, \sigma)_{|\theta_{02}}$ (Supplemental Material \cite{SI}). Figure \ref{fig:fig_4}a shows that $\langle \tau_\textrm{f}(\mu)\rangle$ should be minimized when $\alpha$ is chosen so that $\mu(\alpha) \approx \phi_{\rm u}$. Our numerical results indeed confirm that, for a given value of $\alpha$, the means of the distributions of the angles in the upper level after to lower has closed completely (exemplary distributions shown in Fig. \ref{fig:fig_4}b) as a function of the lower level’s target angle $\phi_{\rm l}$ closely resemble the upper level’s target angles $\phi_{\rm u}$ of the structures with the highest folding rate at that $\alpha$ (Figs. 4c and S5), thus effectively providing a design rule for the choice of the optimal coupling parameter $\alpha$ between levels based on the target geometry.

In conclusion, we have shown how the folding of microscopic hierarchical two-level kirigami structures in a fluid develops a natural degree of coupling between levels due to hydrodynamic resistance. This coupling is an essential factor to account for when it comes to understanding the folding dynamics of similar structures as both folding rate and yield are impacted in its absence. Moreover, we show how rational design rules (e.g. based on solving a first passage time problem) can be used to explain the emergence of an optimal coupling parameter for maximizing folding rates and yield as a function of the geometry of the target structure. In the future, we envisage that these design rules could be adopted to engineer hydrodynamic resistances and active hinges for designing fast self-folding hierarchical multi-level kirigami structures for applications in, e.g., soft robotics \cite{Wang2018,Rus2018,Miskin2020,Bacchetti2022,dong2022} and mechanical actuators and metamaterials \cite{Malachowski2014,Bertoldi2017,Fang2018,Niu2019,Zhang2022}. 

{\bf Acknowledgements.} MPB, QXP and GV are grateful to the studentship funded by the A*STAR-UCL Research Attachment Programme through the EPSRC M3S CDT (EP/L015862/1). NAMA acknowledges financial support from the Portuguese Foundation for Science and Technology (FCT) under the contracts no. UIDB/00618/2020 and UIDP/00618/2020. BJW is supported by the Royal Commission for the Exhibition of 1851. NAMA and GV acknowledge support from the UCL MAPS Faculty Visiting Fellowship programme.

{\bf Author Contributions.} Author contributions are defined based on the CRediT (Contributor Roles Taxonomy) and listed alphabetically. Conceptualization: NAMA, GV. Data Curation: MPB, NAMA, BJW, DRH, GV. Formal analysis: MPB, NAMA, BJW, DRH, GV. Funding acquisition: QXP, RN, GV. Investigation: MPB, NAMA, BJW, DRH, GV. Methodology: MPB, NAMA, BJW, DRH, GV. Project administration: QXP, RN, GV. Resources: QXP, RN, GV. Software: MPB. Supervision: NAMA, QXP, RN, GV. Validation: MPB, NAMA, GV. Visualization: MPB, BJW. Writing – original draft: MPB, NAMA, BJW, DRH, GV. Writing – review and editing: MPB, NAMA, BJW, DRH, QXP, RN, GV.

\renewcommand{\thefigure}{S\arabic{figure}}
\setcounter{figure}{0}

\renewcommand{\vec}[1]{\bm{#1}}
\newcommand{\abs}[1]{\left\lvert#1\right\rvert}
\newcommand{\intd}[1]{~\mathrm{d}#1}
\newcommand{\cross}{\mathop{\times}}

\newcommand{\angnear}{\theta_{\textrm{l}}}
\newcommand{\angdiff}{\theta_{\textrm{u}}}
\newcommand{\angfar}{\varphi}
\newcommand{\tangent}{\vec{t}}
\newcommand{\normal}{\vec{n}}
\newcommand{\tnear}{\vec{t}_{\textrm{l}}}
\newcommand{\tfar}{\vec{t}_{\textrm{u}}}
\newcommand{\nnear}{\vec{n}_{\textrm{l}}}
\newcommand{\nfar}{\vec{n}_{\textrm{u}}}
\newcommand{\x}{\vec{x}}

\newcommand{\ex}{\vec{e}_x}
\newcommand{\ey}{\vec{e}_y}
\newcommand{\ez}{\vec{e}_z}

\newcommand{\f}{\vec{f}}
\newcommand{\CN}{C^n}
\newcommand{\CT}{C^t}
\newcommand{\CNl}{C^n_{\textrm{l}}}
\newcommand{\CTl}{C^t_{\textrm{l}}}
\newcommand{\CNu}{C^n_{\textrm{u}}}
\newcommand{\CTu}{C^t_{\textrm{u}}}

\newcommand{\Fnear}{F_{\textrm{l}}}
\newcommand{\Ffar}{F_{\textrm{u}}}
\newcommand{\hl}{h_{\textrm{l}}}
\newcommand{\hu}{h_{\textrm{u}}}
\newcommand{\m}{m}

\section*{\label{sec:mathsdesc}Hydrodynamic coupling in the overdamped motion of two hinged sheets}

Here, we formulate a relatively simple mechanical model for the motion of two flat sheets that are connected by a hinge, evolving in an ambient fluid on such a scale that inertial effects are negligible. Seeking a first-order approximation and analytical tractability, we make use of \emph{resistive force theory} \cite{Hancock1953, Gray1955} to couple the object's motion to the fluid. This approach neglects secondary hydrodynamic interactions between the sheets and assigns constant drag coefficients to each sheet. 

\begin{figure}[h!]
    \centering
    \includegraphics[width=9cm]{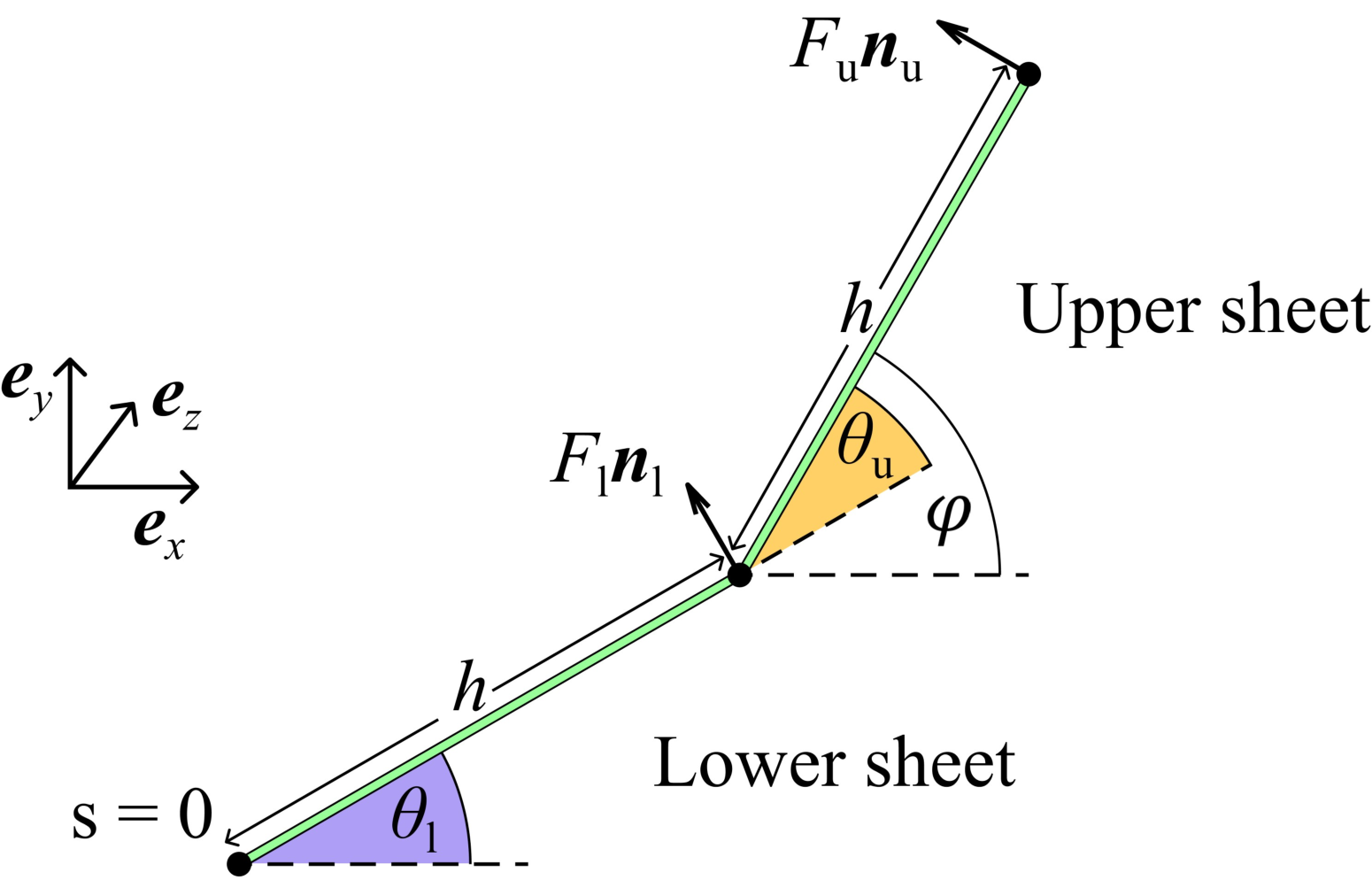}
    \caption{\textbf{Model parameterization.} Each set of the kirigami's lateral faces (Fig. 1) can be modeled as two flat sheets connected by a hinge with the lower sheet tethered to the substrate ($s = 0$). In the schematics, the sheets are assumed to extend in and out of the plane of the diagram, with all motion occurring within the plane. The configuration of the connected object is captured by the orientation $\theta_{\rm l}$ of the lower sheet and $\angfar$ of the upper sheet (both defined in the laboratory frame, $\{\ex,\ey,\ez\}$). Alternatively, the orientation of the upper sheet can be captured by $\angdiff \coloneqq \angfar - \angnear$. The object is driven by two point forces, $F_{\rm l}$ and $F_{\rm u}$, acting respectively on the lower and upper hinges and directed along the normals to each sheet ($\bm{n}_{\rm l}$ and $\bm{n}_{\rm u}$). 
    }
    \label{fig:figset}
\end{figure}

We focus on the two-dimensional motion of the sheets in a plane orthogonal to both, thus constraining their motion as shown in Fig. \ref{fig:figset}. In this set up, the two flat sheets are assumed to extend out of the plane of the figure. For simplicity, we consider identical sheets of equal length, $h$, here, although we quote more general results for sheets of different length below.  Pinning the end of the lower sheet to a substrate ($s = 0$), the configuration of the connected object in this setting is captured by the orientation $\angnear$ of the lower sheet and the orientation $\angfar$ of the upper sheet, both measured relative to an axis fixed in the laboratory frame. We will first formulate the model in terms of $\angnear$ and $\angfar$ and later rewrite it in terms of $\angnear$ and $\angdiff \coloneqq \angfar - \angnear$, the relative angle between the sheets. Noting that each sheet is assumed to be flat, we define the tangents $\tnear$ and $\tfar$ and normals $\nnear$ and $\nfar$ to each sheet, given respectively by
\begin{subequations}
\begin{align}
    \tnear &= \cos{\angnear}\ex + \sin{\angnear}\ey\,, & \nnear &= -\sin{\angnear}\ex + \cos{\angnear}\ey\,,\\
    \tfar &= \cos{\angfar}\ex + \sin{\angfar}\ey\,, & \nfar &= -\sin{\angfar}\ex + \cos{\angfar}\ey\,,
\end{align}
\end{subequations}
where $\ex$ and $\ey$ form part of the right-handed orthonormal basis $\{\ex,\ey,\ez\}$ of the laboratory frame (Fig. \ref{fig:figset}). With this notation, we can write the position $\x(s)$ of a material point in the plane as
\begin{equation}
    \x(s) = \begin{cases}
        s\tnear & \text{if $s\in[0,h)$}\,,\\
        h\tnear + (s-h)\tfar & \text{if $s\in[h,2h]$}\,,
    \end{cases}
\end{equation}
where $s\in[0,2h]$ is an arclength parameter measured from the pinned end, which we have taken to be the origin of the laboratory frame ($s=0$, Fig. \ref{fig:figset}). We will relate the forces imparted on each sheet by the fluid to the velocity of the sheets, which we compute as
\begin{equation}\label{eq: velocity}
    \dot{\x}(s) = \begin{cases}
        s\dot{\angnear}\nnear & \text{if $s\in[0,h)$}\,,\\
        h\dot{\angnear}\nnear + (s-h)\dot{\angfar}\nfar & \text{if $s\in[h,2h]$}\,.
    \end{cases}
\end{equation} \par
Employing resistive force theory \cite{Hancock1953, Gray1955}, we suppose the existence of tangential and normal resistive force coefficients, $\CT$ and $\CN$, for each sheet such that the force per unit length $\f(s)$ exerted on each sheet is given by
\begin{equation}\label{eq: RFT}
    \f(s) = \CT\left[\dot{\x}(s)\cdot\tangent(s)\right]\tangent(s) + \CN\left[\dot{\x}(s)\cdot\normal(s)\right]\normal(s)\,,
\end{equation}
where $\tangent(s)$ and $\normal(s)$ denote the tangent and normal associated with arclength $s$, for brevity. Here, $\CT<0$ and $\CN<0$, so that $\f$ indeed captures a notion of hydrodynamic drag. Again, we assume for simplicity that each sheet has the same resistive force coefficients; this assumption is also relaxed in the general results quoted below. In fact, the results of our analysis will not depend strongly on the appropriate values of $\CN$ and $\CT$, though we remark that, in the context of thin filaments rather than sheets, the approximate relation $\CN \approx 2\CT$ holds \cite{Gray1955,Hancock1953}, with $\CT/\CN < 1$ expected for more general sheet shapes \cite{Koens2016}. \par
In the inertialess regime appropriate for the scales involved in this problem, equations of moment balance apply on each sheet. That is, the moments induced by the hydrodynamic drag must instantaneously balance those induced by any applied forces or torques. These two conditions will uniquely determine $\angnear$ and $\angfar$. Here, we will assume that the system is driven by two point forces that act at the connecting hinge and the free end of the upper sheet, denoting the magnitudes of these forces by $\Fnear$ and $\Ffar$, respectively. Further, we will consider only prescribed forces that drive rotation of the sheets by acting in the normal direction, so that they exert forces $\Fnear\nnear$ and $\Ffar\nfar$, respectively. We additionally allow for the possibility of a prescribed (though potentially state-dependent) point torque $\m\ez$ to act at the connecting hinge. \par
With these driving terms, it is convenient to write the two constraints of moment balance as two integrals, one over only the upper sheet and another over both sheets, resolving moments about the hinge and the lower end, respectively. Explicitly, we have
\begin{subequations}\label{eq: moment balance}
\begin{align}
    0 &= \ez \cdot \int_{h}^{2h}[\x(s) - \x(h)] \cross \f(s)\intd{s} + h\Ffar + \m\,,\\
    0 &= \ez \cdot \int_0^{2h}\x(s) \cross \f(s)\intd{s} + h\Fnear + h\Ffar[1 + \cos{(\angfar - \angnear)}]\,,
\end{align}
\end{subequations}
imposing a moment-free condition at the end of the upper sheet. Computing the required integrals along each sheet yields the explicit linear system of equations
\begin{equation}
    -\frac{\CN h^2}{6}\begin{bmatrix}
        2 & 2+ 3\cos{\angdiff}\\
        2 + 3\cos{\angdiff} & 4 + 6\cos{\angdiff}\left(1 + \cos{\angdiff}\right) + 6\frac{\CT}{\CN}\sin^2{\angdiff}
    \end{bmatrix}
    \begin{bmatrix}
        \dot{\angdiff}\\
        \dot{\angnear}
    \end{bmatrix}
    =
    \begin{bmatrix}
        1 & 0 & 1\\
        (1 + \cos{\angdiff}) & 1 & 0
    \end{bmatrix}
    \begin{bmatrix}
        \Ffar\\
        \Fnear\\
        m/h
    \end{bmatrix}\,,
\end{equation}
recalling that $\angdiff = \angfar - \angnear$ is the relative angle between the two sheets and that $\CN < 0$. We can manipulate this system by subtracting $(1 + \cos\theta_{\rm u})$ multiplied by the first equation from the second to simplify the dependence on $\Ffar$ and $\Fnear$:
\begin{equation}\label{eq: full system force isolated}
    -\frac{\CN h^2}{6}\begin{bmatrix}
        2 & -2 \hat{\alpha}(\angdiff)\\
        \cos{\angdiff} & -\cos{\angdiff}/\hat{\beta}(\angdiff)
    \end{bmatrix}
    \begin{bmatrix}
        \dot{\angdiff}\\
        \dot{\angnear}
    \end{bmatrix}
    =
    \begin{bmatrix}
        1 & 0 & 1\\
        0 & 1 & -(1 + \cos{\angdiff})
    \end{bmatrix}
    \begin{bmatrix}
        \Ffar\\
        \Fnear\\
        m/h
    \end{bmatrix}\,,
\end{equation}
where
\begin{equation}
    \hat{\alpha}(\angdiff) = - \left[ 1  + \frac{3}{2}\cos{\angdiff}\right]\,, \quad 
 \hat{\beta}(\angdiff) = \frac{-\cos{\angdiff}}{ 2 + \cos{\angdiff}\left(1 + 3\cos{\angdiff}\right) + 6(\CT/\CN)\sin^2{\angdiff}}\,.
\end{equation}

\begin{figure}[!]
    \centering
    \includegraphics[width = 8.6cm]{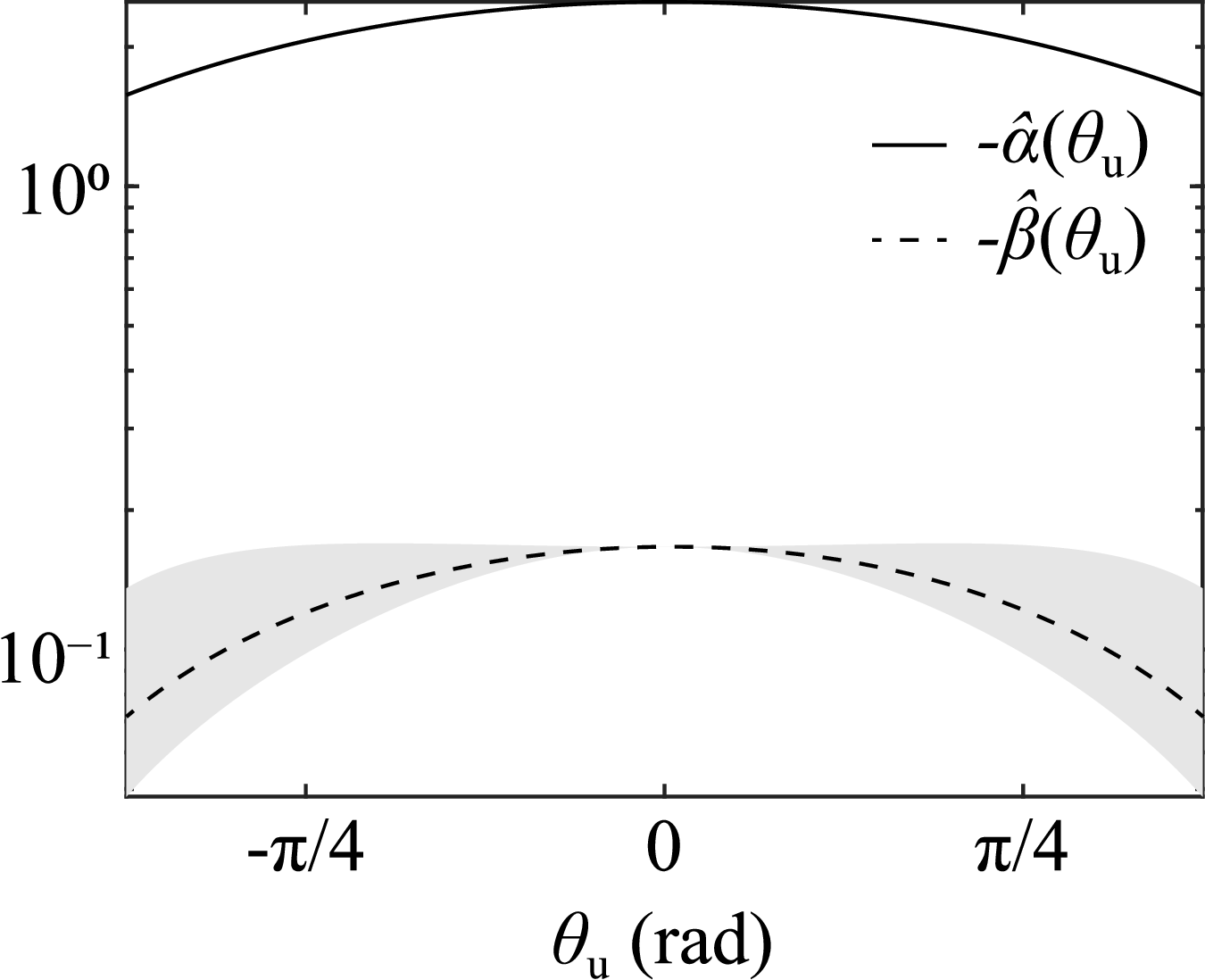}
    \caption{\textbf{Relative coupling between $\angdiff$ and $\angnear$ as a function of $\angdiff$.} Functions $-\hat\alpha(\angdiff)$ (solid) and $-\hat\beta(\angdiff)$ (dashed) on a logarithmic scale for a range of $\angdiff$, for the case of equal sheet lengths and equal resistive force components, highlighting the order-of-magnitude dominance of the former over the latter. Recall that these represent the negative feedback of motion of the lower face on the upper and motion of the upper on the lower, respectively (see eq. (\ref{eq: maintext_form})).  The lines have $\CT/\CN = 1/2$, appropriate for slender filaments; the shaded region shows the range of possible values for $0\leq \CT/\CN \leq 1$, illustrating that the qualitative result is largely insensitive to this ratio.
    }
    \label{fig:my_label}
\end{figure}

Equation (\ref{eq: full system force isolated}) shows how the angular variables relate to the corresponding forces, and highlights how a particular choice of the active hinge torque $\m$ could qualitatively change this relationship (for instance if $m \propto \dot{\theta}_{\rm u}$ or $m \propto \dot{\angnear}$). For simplicity here, we focus on a free hinge with $\m=0$, leading to the results quoted in the main text:
\begin{subequations} \label{eq: maintext_form}
\begin{align}
     \dot{\theta}_{\rm u} &= \hat{\alpha}(\theta_{\rm u})\dot{\theta}_{\rm l} -\frac{3}{\CN h^2} F_{\rm u},\\
    \dot{\theta}_{\rm l} &= \hat{\beta}(\theta_{\rm u})\dot{\theta}_{\rm u} + \frac{6\hat{\beta}(\theta_{\rm u})}{\cos(\theta_{\rm u}) \CN h^2} F_{\rm l}.
\end{align}
\end{subequations}
For sheets oriented approximately parallel to one another, in particular those with $\abs{\angdiff}\ll1$, the functions $\hat{\alpha}$ and $\hat{\beta}$ simplify substantially to leading order in powers of $\angdiff$ to yield the approximate, constant-coefficient system from Eq. (\ref{eq: full system force isolated}),
\begin{subequations}
\begin{align}
    2\dot{\angdiff} &= - 5\dot{\angnear} - \frac{6}{\CN h^2}\Ffar\,,\\
    \dot{\angdiff} &= - 6\dot{\angnear} - \frac{6}{\CN h^2}\Fnear\,.
\end{align}
\end{subequations}
From this, we can see that if one were to drive the system via only the lower sheet, so that $\Ffar = 0$ and $\Fnear\neq0$, the reorientation rate of the upper sheet is $\dot{\angdiff} \sim -5 \dot{\angnear} /2$. Conversely, if one were to drive the system via only the upper sheet, so that $\Fnear=0$ and $\Ffar\neq0$, the reorientation rate of the lower sheet is $\dot{\angnear} \sim -\dot{\angdiff}/6$. Hence, in this sense, the coupling between $\angdiff$ and $\angnear$ is dominated by a negative feedback in the motion of the upper sheet. More generally, the application of any driving force of the form considered here has a greater impact on the relative orientation of the upper sheet than it does on the orientation of the lower sheet, at least in the small relative-angle regime. The validity of these conclusions outside this small-angle regime hinges on the relative size of $\hat{\alpha}$ and $\hat{\beta}$ in Eq. \ref{eq: maintext_form}. These are shown for a range of angles in Fig. \ref{fig:my_label}, which illustrates that they differ by approximately an order of magnitude across a large range of $\angdiff$; this difference confirms that forcing drives larger changes in $\angdiff$ than in $\angnear$ in general.

More generally, it is straightforward, if somewhat involved, to repeat this analysis allowing for different lengths $\hl$ and $\hu$ of each sheet, and different resistive force coefficients $\CNu$, $\CNl$, $\CTu$, $\CTl$. Omitting the details and no longer assuming $m=0$, the equivalent expressions in that case also involve the length ratio $R \equiv \hu/\hl$ and the normal resistive force ratio $\Gamma \equiv \CNu/\CNl$ and are
\begin{subequations}\label{eq: everything}
\begin{align}
  \dot{\theta}_{\rm u} &= \hat{\alpha}(\theta_{\rm u})\dot{\theta}_{\rm l} -\frac{3}{\CNu \hu^2} \left(F_{\rm u} + \frac{\m}{\hu}\right),\\
    \dot{\theta}_{\rm l} &= \hat{\beta}(\theta_{\rm u})\dot{\theta}_{\rm u} + \frac{6\hat{\beta}(\theta_{\rm u})}{\CNu \hu^2 (3-2R)\cos{\theta_{\rm u}}} \left[ F_{\rm l} - R (1+\cos{\theta_{\rm u}}) \frac{m}{\hu}\right]\,,
\end{align}
\end{subequations}
with
\begin{equation}
    \hat{\alpha}(\angdiff) = - \left[ 1  + \frac{3}{2R}\cos{\angdiff}\right]\,, \quad 
 \hat{\beta}(\angdiff) = \frac{(3-2R)\cos{\angdiff}}{2/(\Gamma R^2) + \cos{\angdiff}\left[3 - 2 R + \left(6/R - 3\right)\cos{\angdiff}\right] + 6[\CTu/(\CNu R)]\sin^2{\angdiff}}\,.
\end{equation}
These expressions illustrate how the strength of coupling between the sheets can be modified by suitable choice of the ratios $R$ and $\Gamma$, as well as by the active hinge torque $m$ if it has some dependence on $\dot{\theta}_{\rm u}$ or $\dot{\theta}_{\rm l}$. In particular, these expressions provide motivation for exploring a range of values - including both positive and negative - for the simplified constant coupling parameter $\alpha$ in the main text. 

\section*{\label{sec:diffusioncoeef}Deriving a diffusion coefficient}

We derive an approximate rotational diffusion coefficient for each face of the kirigami, assuming that faces are entirely uncoupled. Let us consider a face in a fluid at thermostat temperature $T$, connected by a hinge to a surface such that it can only rotate around one edge. If this edge is a principle axis of inertia, the equation of motion for the angle $\theta(t)$ with the surface is,
\begin{equation}
    I\ddot{\theta}(t)=-f_{\rm r}\dot{\theta}(t)+\Xi\xi(t) \, ,
\end{equation}
\noindent where $t$ is time, $I$ is the corresponding momentum of inertia, $f_{\rm r}$ is the frictional resistance coefficient, $\xi(t)$ is a stochastic term with zero mean and variance one, and $\Xi$ sets the amplitude of the thermal fluctuations in the force. Thus, for $\theta(0)\equiv 0$, the mean-square displacement of the angle $\theta(t)$ is,
\begin{equation}\label{eq:sup.theta2}
\langle\theta^2(t)\rangle\sim\left[\frac{\Xi^2}{If_{\rm r}^2}\right]t \, .
\end{equation}
\noindent The dependence of $\Xi$ on $T$ and $f_r$ can be derived by invoking equipartion as follows. The equation for the angular velocity $\omega\equiv\dot{\theta}$ is
\begin{equation}
    I\dot{\omega}(t)=f_{\rm r}\omega(t)+\Xi\xi(t) \, .
\end{equation}
Assuming the same initial velocity $\omega(0)\equiv\omega_0$, the mean-square velocity is then,
\begin{equation}
    \langle\omega^2(t)\rangle=\omega_0^2\exp\left(-\frac{2f_{\rm r}}{I}t\right)+\frac{\Xi^2}{2If_{\rm r}}\left[1-\exp\left(-\frac{2f_{\rm r}}{I}t\right)\right]  \, .
\end{equation}
Asymptotically, from the equipartition theorem, $\omega^2=k_{\rm B}T/I$, where $k_{\rm B}$ is the Boltzmann constant, and thus,
\begin{equation}
    \Xi=\sqrt{2f_{\rm r}k_{\rm B}T} \ \ .
\end{equation}
Combining with Eq.~\eqref{eq:sup.theta2}, we obtain the Einstein relation,
\begin{equation}
    D = \frac{k_{\rm B} T}{f_{\rm r}} \ \ .
\end{equation}

The frictional resistance coefficient $f_{\rm r}$ of the plate is computed as the ratio of applied torque and the resulting angular velocity. Computing this ratio is in fact a subproblem of the above section, exactly equivalent to considering the torque balance on only the bottom face. This immediately yields
\begin{equation}
    f_{\rm r} = -\frac{\CN h^3}{3}
\end{equation}
for normal resistive coefficient $\CN$. In the absence of an exact expression for $\CN$ for a general plate, we approximate the face as a circular disc of radius $h/\sqrt{\pi}$. A circular disc with this radius that moves at unit speed normal to its radii in a medium of viscosity $\mu$ experiences total drag $16\mu h/\sqrt{\pi}$ \cite{Davis1991a}. Crudely dividing this quantity by its diameter to yield a representative drag per unit length gives a drag coefficient of $\CN = -8\mu$. Hence, we approximate
\begin{equation}
    f_{\rm r} \approx \frac{8\mu h^3}{3}\,.
\end{equation}
Finally, the Einstein relation gives
\begin{equation}
    D = \frac{k_{\rm B} T}{f_{\rm r}} = \frac{3k_{\rm B} T}{8\mu h^3}
\end{equation}
as an estimate for the rotational diffusion coefficient of a single face.

\begin{figure}[!]
    \centering
    \includegraphics[width=15cm]{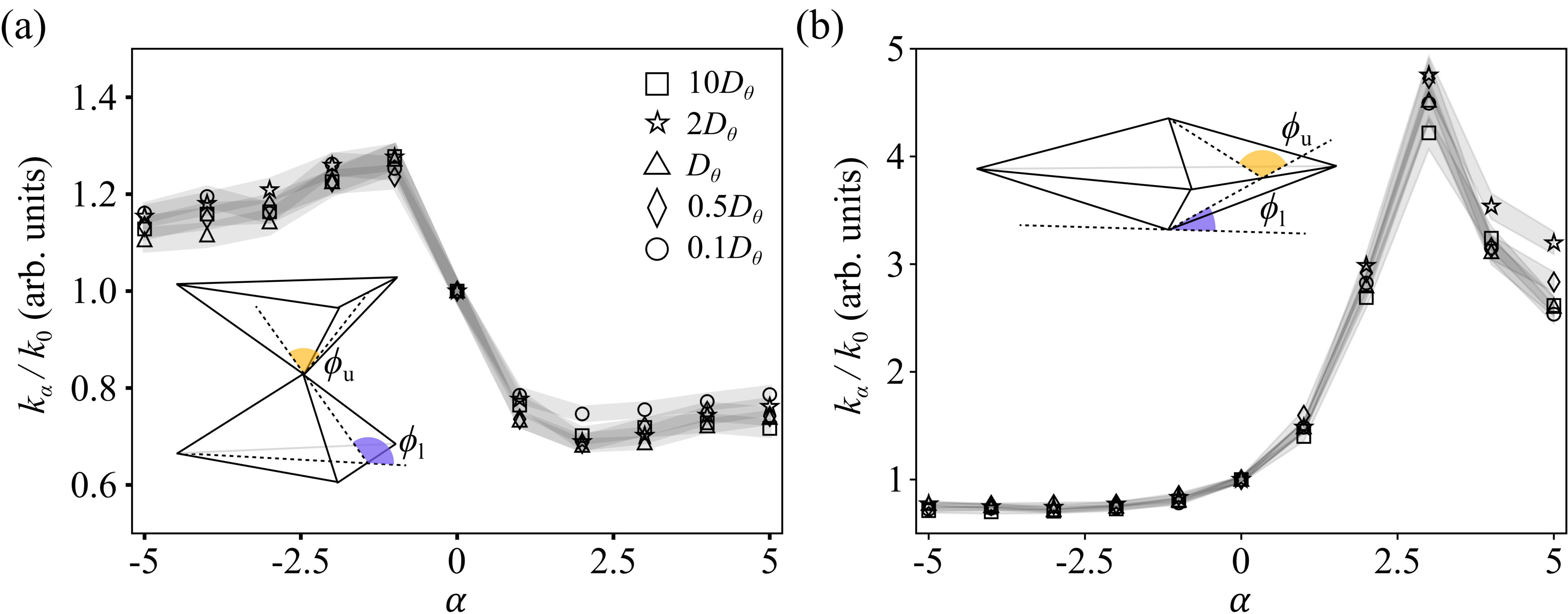}
    \caption{\label{fig:dif}\textbf{Dependence of the normalized folding rate on the rotational diffusion coefficient.} Normalized folding rate  $k_{\alpha}/k_0$ as a function of coupling parameter $\alpha$ for two exemplary kirigami structures (as in Fig. 2) with different rotational diffusion coefficients: (a) a hourglass ($\phi_\textrm{l} = \frac{\pi}{3} \, {\rm rad}$, $\phi_\textrm{u} = -\frac{\pi}{6} \, {\rm rad}$) and (b) a diamond ($\phi_{\textrm{l}} = 0.61$ rad, $\phi_{\textrm{u}} = 1.92$ rad). Folding rates are normalized to $k_{0}$ ($k_{\alpha}$ for $\alpha = 0$). The shaded areas represent one standard error around the average values from $5000$ folding events per value of rotational diffusion constant, each lasting up to the cutoff time $\tau_{\rm cut} = 2 \cdot 10^8 \Delta \tau$ (with $\Delta \tau$ being the simulation time step). 
    }
\end{figure}

In our simulations, the rotational diffusion coefficient was fixed to $D = 0.64 \ {\rm rad^2} \ {\rm s^{-1}}$. Figure \ref{fig:dif} shows that, qualitatively, the dependence (i.e. the position of the peak) of the normalized folding rate $k_\alpha/k_0$ ($k_0$ being $k_\alpha$ for $\alpha =0$) on $\alpha$ is largely independent of the exact value of the diffusion coefficient for structures whose face height $h$ is on the scale of $\mathcal{O}({\rm \mu m})$ to $\mathcal{O}( 10 \, {\rm \mu m})$.  
At larger scales (i.e. by increasing $h$ further), gravitational torques will also need to be accounted for to determine the overall folding behavior \cite{Leong2007}. 

\section*{\label{sec:analytical}Mapping to a first passage time problem}

For pyramids with less than five faces, it has been shown that the folding time is dominated by the time it takes for the first pair of faces to close \cite{Simoes2021,melo2020}. Equally, for two-level structures, if we assume that the lower level folds much faster than the upper one, the folding time can be estimated as the time it takes for the first pair of faces of the upper level to meet at their target angle $\phi_{\rm u}$, after the lower level has closed completely. The time evolution of the angles of these two faces can be mapped into a two-dimensional Brownian walk (one dimension for each face) and we can neglect collisions between faces, as suggested by molecular dynamics simulations in Ref.~\cite{melo2020}. If we discretize the configuration space, the motion of the faces corresponds to a random walk on a lattice with $L$ points per dimension with reflective boundaries. In the most general case, such a two-dimensional lattice is bound at $-\pi$ and $\pi$ along each direction and spaced every $\Delta = \pi/180$.  
Finding the closing time of a pair of faces corresponds to determining the first-passage time for the two-dimensional random walk visiting the lattice site $\pmb{\phi} = (\phi_{\rm u}, \phi_{\rm u})$, equivalent to both faces reaching $\phi_{\rm u}$ at the same time. The walk begins at $\bm{\theta}_0 = (\theta_{01}, \theta_{02})$, the subscripts $1$ and $2$ referring to each of the two faces respectively. The Mean First Passage Time (MFPT) for such a walk can be obtained through \cite{Giuggioli2020}
\begin{equation}\label{eq: mfpt}
    \begin{split}
        T_{\bm{\theta}_0 \rightarrow \pmb{\phi}} = 8\Biggr[\sum_{\theta_1 = 1, \theta_2 = 1}^{L-1} K_{\theta_1, \theta_2}(\bm{\theta}_0,\pmb{\phi}) + \frac{1}{2}\sum_{\theta_1 = 1}^{L-1} K_{\theta_1, 0}(\bm{\theta}_0,\pmb{\phi}) + \frac{1}{2}\sum_{\theta_2 = 1}^{L-1} K_{0, \theta_2}(\bm{\theta}_0,\pmb{\phi})\Biggr] \, ,
    \end{split}
\end{equation}
where $\bm{\theta} = (\theta_{1}, \theta_{2})$ represents a generic lattice position, and
\begin{equation}\label{eq:k} 
    \begin{split}
        K_{\theta_1, \theta_2}(\bm{\theta}_0,\pmb{\phi}) = \Biggr[{\rm cos}^2\biggr[\Big(\phi_{\rm u}+\frac{1}{2}\Big)\frac{\theta_1\pi}{L}\biggr]{\rm cos}^2\biggr[\Big(\phi_{\rm u}+\frac{1}{2}\Big)\frac{\theta_2\pi}{L}\biggr] 
     - {\rm cos}\biggr[\Big(\theta_1+\frac{1}{2}\Big)\frac{\theta_1\pi}{L}\biggr]{\rm cos}\biggr[\Big(\theta_2+\frac{1}{2}\Big)\frac{\theta_2\pi}{L}\biggr] \times \\
    \times {\rm cos}\biggr[\Big(\phi_{\rm u}+\frac{1}{2}\Big)\frac{\theta_1\pi}{L}\biggr]
    {\rm cos}\biggr[\Big(\phi_{\rm u}+\frac{1}{2}\Big)\frac{\theta_2\pi}{L}\biggr]\Biggr]
    \times \Biggr[2 - {\rm cos}\Big(\frac{\theta_1\pi}{L}\Big) - {\rm cos}\Big(\frac{\theta_2\pi}{L}\Big)\Biggr]^{-1} \, .
    \end{split}
\end{equation}

Equation (\ref{eq: mfpt}) represents the MFPT while Eq. (\ref{eq:k}) is the lattice propagator function for a two-dimensional random walk in a bounded domain \cite{Giuggioli2020}. A summation over the whole discretized space is therefore considered in order to establish the possible evolutions of the two-dimensional walk starting at $\bm{\theta}_0$ and ending at $\pmb{\phi}$. 

If we assume that the motion of the different faces in the upper level is uncorrelated and that the probability distribution of the angles $\mathcal{N}(\mu, \sigma)$ is a Gaussian reflected at the lattice boundaries, the probability of starting at a given lattice position can be calculated as
\begin{equation}
    P(\bm{\theta}_0) = \mathcal{N}(\mu, \sigma)_{|\theta_{01}}\mathcal{N}(\mu, \sigma)_{|\theta_{02}} \ .
\end{equation}

We can thus obtain an expected MFPT, $\langle\tau_\textrm{f}(\mu)\rangle$, by first calculating $T_{\bm{\theta}_0 \rightarrow \bm{\phi}}$ for every lattice position, multiplying it by the probability $P(\bm{\theta}_0)$ of starting at that point and integrating over the whole range of possible starting positions $\bm{\theta}_0$
\begin{equation}
    \langle \tau_\textrm{f}(\mu)\rangle = \iint T_{\bm{\theta}_0 \rightarrow \bm{\phi}} P(\bm{\theta}_0) d\bm{\theta}_0 \ .
\end{equation}

\clearpage

\section*{\label{sec:figures}Supplemental figures}

\begin{figure}[!]
    \centering
    \includegraphics[width=15cm]{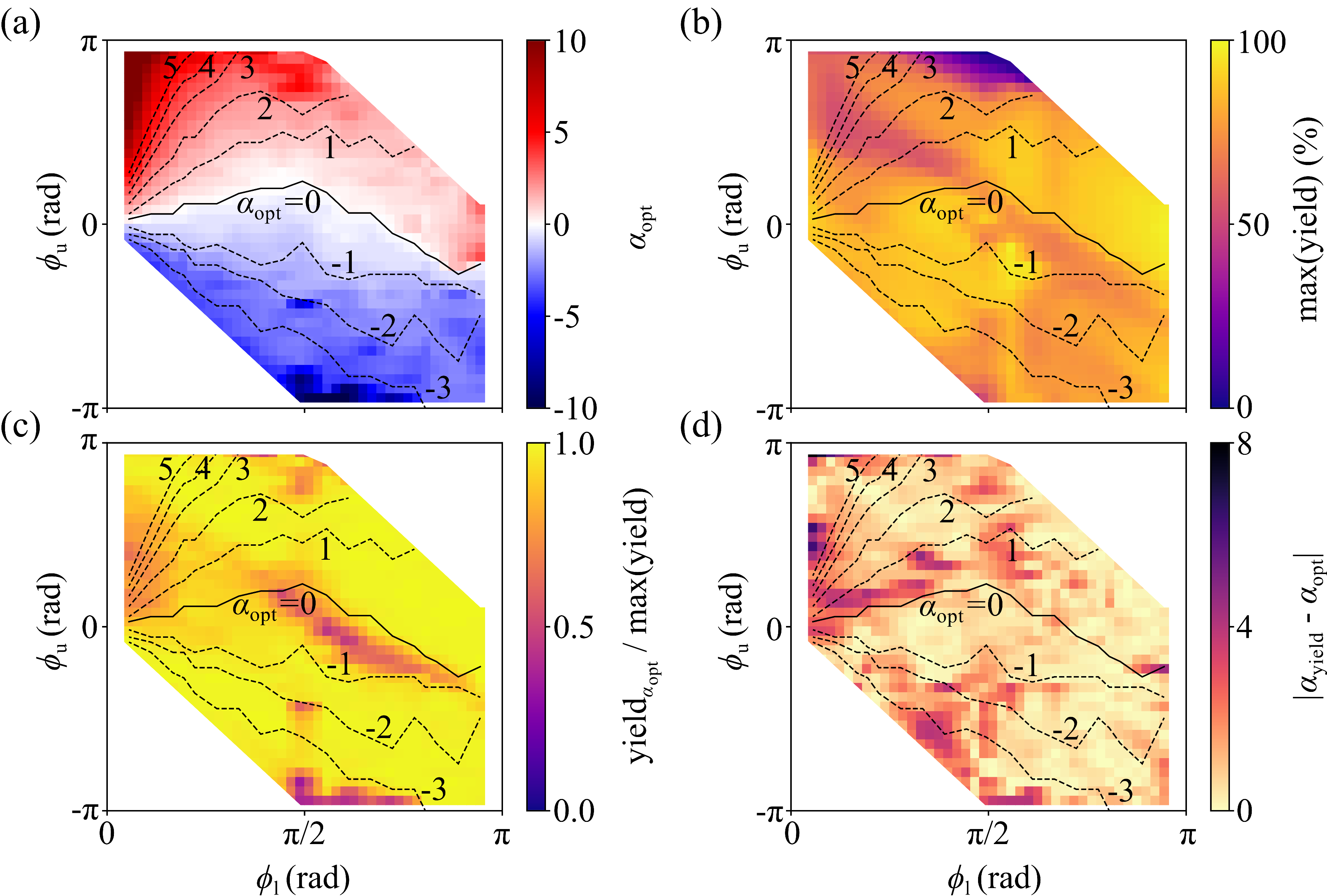}
    \caption{\label{fig:yield}\textbf{Optimal rate vs. optimal yield.} (a) Optimal $\alpha$ value ($\alpha_{\rm opt}$) leading to the highest folding rate ${\rm max}(k_{\alpha})$,
    (b) highest folding yield ${\rm max}({\rm yield})$ at any $\alpha$, (c) ratio between yield at $\alpha_{\rm opt}$ (${\rm yield}_{\alpha_{\rm opt}}$) and ${\rm max}({\rm yield})$ at any value of $\alpha$, (d) distance between $\alpha$ values that optimizes yield ($\alpha_{\rm yield}$) and $\alpha_{\rm opt}$ for different target structures.
    In (a-d), the black dashed isolines show the optimal values of $\alpha$ that lead to the highest folding rates and the black solid lines highlight structures whose folding is optimal in the absence of coupling ($\alpha = 0$) for reference. Each data point is obtained as an average of 5000 folding events. 
    }
    \label{fig:surfaces}
\end{figure}

\begin{figure}[!]
    \centering
    \includegraphics[width=15cm]{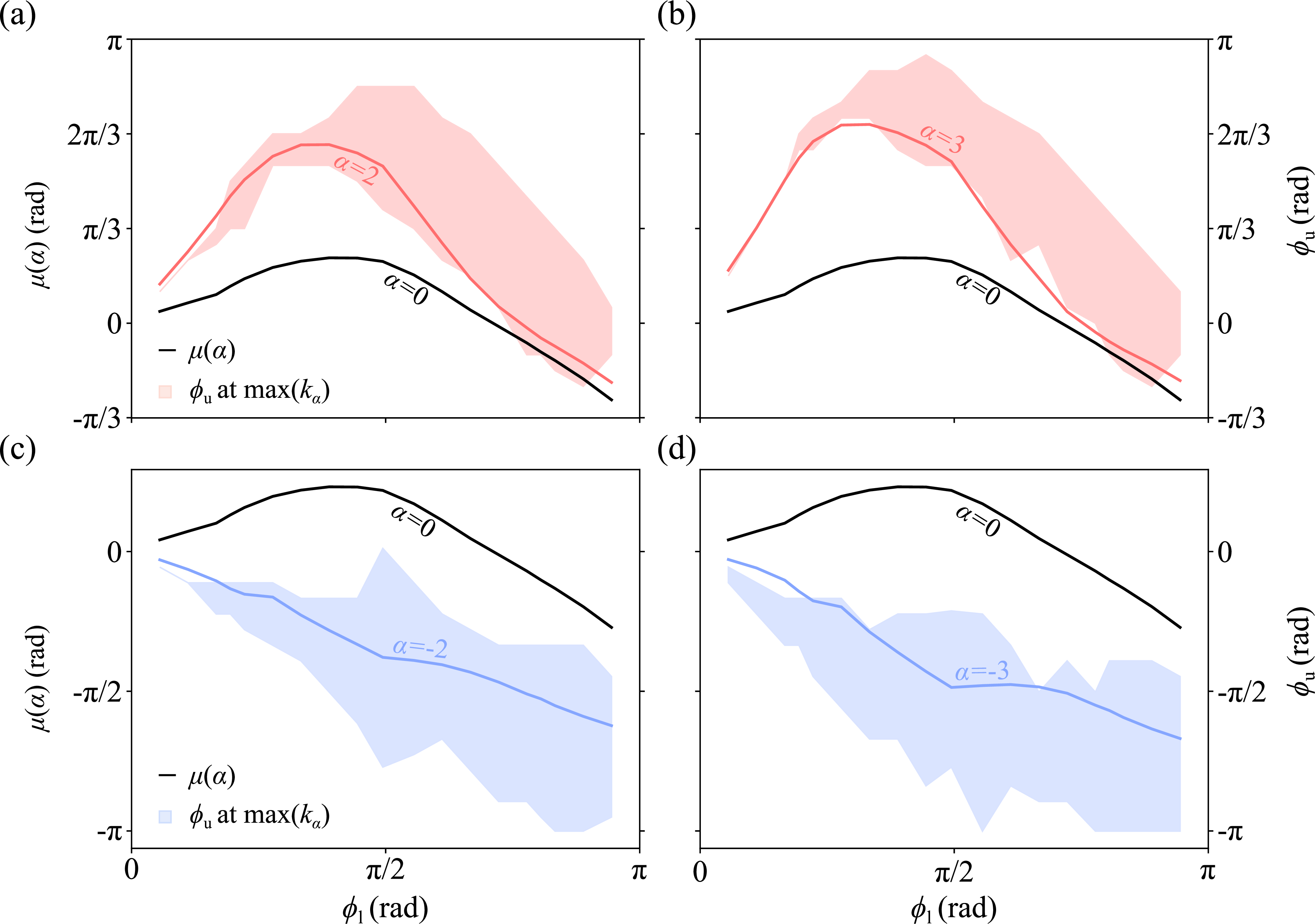}
    \caption{\label{fig:alphas}\textbf{Target geometry and optimal $\alpha$.} 
    For a given $\alpha$, the means $\mu(\alpha) = \langle \theta^{i}_{\rm u}(t_2) \rangle_\alpha$ (solid lines) of the probability distributions of the angles $\theta^i_{\rm u}$ of the upper level's faces at the closing time $t_2$ of the lower level  are well reproduced by the structures (represented as combinations of the two target angles $\phi_{\rm u}$ and $\phi_{\rm l}$) with the highest folding rate at that value of $\alpha$ (shaded regions): (a) $\alpha = 2$, (b) $\alpha = 3$, (c) $\alpha = -2$ and (d) $\alpha = -3$. The black solid lines show $\mu(\alpha=0)$ for reference. The shaded regions represent all structures that could have the highest folding rate at that value of $\alpha$ once the numerical uncertainty is accounted for.  
    For a given structure, this uncertainty around the position of ${\rm max}(k_\alpha)$ is calculated taking the largest between the discretization step in $\alpha$ and the range of data points around the peak falling within the standard error.   
}
    \label{fig:surfaces}
\end{figure}


%

\end{document}